# Temporal dependence of optically-induced photophoretic force on absorbing airborne particles by a power-modulated laser

Gui-hua Chen[1], Lin He[1], Mu-ying Wu[1], and Yong-qing Li[1,2*]

[1]*School of Electronic Engineering & Intelligentization, Dongguan University of Technology, Dongguan, Guangdong 523808, P.R. China*
[2]*Department of Physics, East Carolina University, Greenville, North Carolina 27858-4353, USA*

Photophoretic force due to the optically-induced thermal effect provides an effective way to manipulate the light-absorbing particles suspended in ambient gases. However, how this force temporally responds to the intensity modulation of the illumination light is unclear. Here, by vertically trapping a micron-sized absorbing particle with a negative photophoretic force generated by a focused Gaussian beam, we demonstrate that the temporal change in the photophoretic force in response to the intensity modulation is remarkably slow (with a time constant up to ~1 s) due to the slow change in the particle's temperature. When the trapping beam is turned off for a few tens or hundreds of milliseconds, the trapped particle is found to be pulled up towards the light source by the remained photophoretic force, whereas when the intensity of the trapping beam is increased for a short duration, the particle is pushed off by the radiation pressure. The instantaneous position of the trapped particles following the intensity modulation of the trapping laser is tracked and theoretically modeled. The understanding of the temporal behavior of the photophoretic force would be useful for the control of photophoretic-based optical pulling, transportation, and manipulation of atmospheric particles.

## I. INTRODUCTION

When objects suspended in a gaseous environment are illuminated by a light beam, two types of optically-induced forces are generated. One is optical force, also called radiation pressure force arising from direct momentum transfer between photons and objects due to the absorption, reflection or scattering of the incident light, which has been widely applied for trapping, manipulation, and tractor of physical and biological particles [1-8]. The other is photophoretic force, originating from momentum transfer between the surrounding gas molecules and hotter objects [9-13], which can be several orders of magnitude larger than the radiation force and gravitation force [14,15]. Photophoretic force has recently been applied for manipulating absorbing aerosol particles [15-18], optical pulling or pushing of airborne particles [19-22], and characterization of trapped aerosol particles [23-25]. Ultra-long distance laser pushing and pulling of absorbing particles have been demonstrated over one meter in air [14,20]. The complete control of photophoretic-based optical pulling, transportation and manipulation of atmospheric particles relies on the understanding of temporal properties of the force.

Comparing to the radiation pressure force $F_{rp}$, the temporal dependence of photophoretic force $F_{pp}$ on the intensity change of illumination light is less understood [9,10]. Radiation pressure can be expressed as $F_{rp}=I(1+R)/c$, with $I$ is flux density, $c$ the speed of light, and $R$ the reflectivity [11]. When illumination light is changed rapidly, $F_{rp}$ changes immediately following the change in photon flux of the incident beam [1,2]. However, the temporal change in the photophoretic force $F_{pp}$ is unclear, because the change in particle's temperature could be much slower than the change in light intensity, whereas the temperature difference between the heated particles and surrounding molecules results in $F_{pp}$ [9-11]. Based on the variation in the surface temperature or in the accommodation coefficient of the particles, two types of photophoretic forces, $F_{\Delta T}$ and $F_{\Delta \alpha}$, can be induced [9,10]. The first type of photophoretic force $F_{\Delta T}$ is caused by inhomogeneous heating of the particle, leading to $F_{\Delta T}$ pointing from the hot to the cold side and always parallel to the light propagation direction. For weakly absorbing particles, a negative $F_{\Delta T}$ force could be induced due to the lens effect [10]. For highly absorbing particles with the size larger than the wavelength, there is more absorption on the illuminated side and, therefore, a positive $F_{\Delta T}$ force is usually induced pointing along the laser beam [10]. The second type of photophoretic force $F_{\Delta \alpha}$ is induced by the difference in heat exchange between the gas molecules and the particle surface even if the particle is heated evenly, leading to a body-fixed $F_{\Delta \alpha}$ force pointing from the side of higher to the side of lower accommodation coefficient [9,10]. It remains unclear to what degree $F_{\Delta \alpha}$ and $F_{\Delta T}$ might contribute, but they can coexist for irregular particles. It was estimated that under atmospheric pressure or below, practically the $F_{\Delta \alpha}$ force is dominant for micron-sized particles with good heat conductivity [9,10], which considerably reduces the temperature difference between the illuminated side and the backside and thus reduces the $F_{\Delta T}$ force. While most previous work used constant light sources to generate $F_{pp}$ under thermal equilibrium condition [15-25], the temporal properties of $F_{pp}$ induced by power-modulated light had not been studied. Recent works suggest that the dynamics induced by time-varying photophoretic potentials is not trivial [22,26,27], and therefore, the understanding of the temporal behavior of the photophoretic force would be important to the emerging topic of particles trapped in time-varying potentials [26,27].

In this work, we experimentally demonstrate that the temporal change in photophoretic force $F_{pp}$ in response to the intensity modulation can be remarkably slow due to the slow change in particle's temperature. We use a downward focused Gaussian beam to trap a highly absorbing particle, which induces a vertical component of $F_{pp}$ to balance the gravitation force $F_G$ and radiation pressure $F_{rp}$. We show that when the trapping beam is turned off for a few tens or hundreds of milliseconds, the particle is found to be pulled up towards the light source by the remained $F_{pp}$. On the other hand, when the light intensity is suddenly increased for a short duration, the particle is pushed downward by rapidly increased radiation pressure. The dynamic motion of the trapped particles exhibits the temporal dependence of photophoretic force $F_{pp}$. Theoretical modeling of the dynamic motion of the trapped particles in response to the power modulation shows that the $F_{pp}$ force changes with a time constant up to ~1 s, much slower than that of radiation pressure.

This paper is organized as follows. In Sec. II, an experimental scheme to measure the time dependence of photophoretic force, the experimental results and discussion are presented. In Sec. III, theoretical estimation of the photophoretic $F_{\Delta T}$ and $F_{\Delta \alpha}$ forces, as well as gravitation force, radiation pressure force, and buoyant force for micron-sized absorbing particles is presented. Based on this, which of the forces ($F_{\Delta T}$ or $F_{\Delta \alpha}$) that are responsible for the particle levitation by the negative photophoretic force is discussed. In Sec. IV, theoretical modeling of experimental results is presented. In Sec. V, the potential applications of the temporal behavior of the photophoretic force for characterizing some physical properties of the trapped particles by measuring their dynamic motions are discussed. Section VI concludes the paper.

## II. EXPERIMENTAL RESULTS AND DISCUSSION

The experimental scheme to measure the time dependence of photophoretic force is shown in FIG. 1(a), in which a highly absorbing micron-sized particle is trapped by a downward focused Gaussian beam. To hold the particle in the trapping position, a negative $F_{pp}$ force must be induced by the downward laser beam such that $F_{pp}=F_G+F_{rp}$, where $F_{pp}$ is the vector sum of $F_{\Delta \alpha}$ and $F_{\Delta T}$ along z axis (with $\vec{F}_{pp} = \vec{F}_{\Delta \alpha} + \vec{F}_{\Delta T}$). Highly absorbing particles of different sizes were used (FIG. 10 in Appendix A), including carbon spherical powders (size of 2-12 μm, Sigma-Aldrich), graphite powders (size of ~5μm, US Research Nanomaterials), carbon nano-powder clusters (size<50 nm, Sigma-Aldrich), and natural graphite nano-powders (size of 400-1200 nm, US Research Nanomaterials), respectively. It should be noted that glassy carbon spheres have high thermal conductivity ($k_p$=119-165 W·m$^{-1}$·K$^{-1}$) [28] and the magnitude of $F_{\Delta T}$ is estimated to be much smaller than $F_{\Delta \alpha}$ at the atmospheric pressure [see FIG. 5(b) in Sec. III]. The use of large sized particles (>4 μm) ruled out the possibility of the induced negative $F_{pp}$ could be due to the $F_{\Delta T}$ force, because the direction of the $F_{\Delta T}$ force can only point along the direction of light propagation for highly absorbing particles with a size well larger than the wavelength [10]. Therefore, the negative component of $F_{pp}$ is mainly dominant by the $F_{\Delta \alpha}$ force for highly absorbing micron-sized particles, which balances the other forces acting on the particles along beam axis direction. The radiation pressure force is an order of magnitude smaller than $F_G$ and the buoyant force is ~3 orders of magnitude smaller than $F_G$, which can be neglected (see Sec. III). When laser power is fast turned off, $F_{rp}$ drops to zero immediately. If the photophoretic force $F_{pp}$ drops as fast as $F_{rp}$, one would expect that the trapped particle should move down due to the action of $F_G$. However, if $F_{pp}$ decreases slower than $F_{rp}$ and retains for a short period, the trapped particle would move upward first and then drop down under the influence of gravitation force $F_G$. On the other hand, if a positive pulse is applied so that the laser intensity is increased for a short duration, the trapped particle would be pushed down by the increased $F_{rp}$ if $F_{pp}$ does not increase as fast as $F_{rp}$. Therefore, by measuring the dynamic motion of the trapped particle following a pulsed change in laser intensity, it is possible to measure the time dependence of photophoretic force $F_{pp}$ as well as particle's temperature.

The experimental setup is shown in FIG. 1(b). A diode laser (HL6545MG, 660 nm) is focused vertically by a lens ($f$=50 mm) to form the trap [15,23]. The laser beam is linearly polarized in TEM$_{00}$ mode and its power is controlled by the driving current. An absorbing particle in a cuvette is trapped near the focus of the laser beam. The particle powders were placed on the bottom plate of the cuvette together with a few mini steel bars, kept at atmospheric pressure ($p$=760 Torr) and 20 ºC. A magnet was used to steer the steel bars, forcing the carbon particles moving on the bottom plate so that a few particles may be pulled up by the laser beam to the trapping position. The inset in FIG. 1(b) shows the image of a trapped particle (~3 mm below the focal position). The power of diode laser was modulated by a voltage pulse via a field-effect transistor (MOSFET) with a rise or fall time less than 1 μs. A high-speed video camera (~1000 fps) was used to record the scattered light images or bright-field images (illuminated by a light-emitting diode) of the trapped particle, from which particle's positions at different times before and after power modulation can be tracked. The axial speed of the moving particle can be measured by the slope of the position-time graph. A fast photodiode was also used to detect the scattered light of the trapped particle, from which the transverse rotation frequency of the moving particle can be determined [15].

The dependence of the trapping position of a particle in z axis on the incident power $P$ is shown in FIG. 2(a). The result shows that when the incident power is increased, the trapped particle is pushed forward, and when the incident power is decreased, the trapped particle is pulled towards the focal point. The experimental data of the displacement ($\Delta z = z - z_1$) of the trapping position $z$ of a particle is shown in FIG. 2(a), in which $z_1$ is the location where the particle is trapped when the laser power is $P_1$ (=30.2 mW).

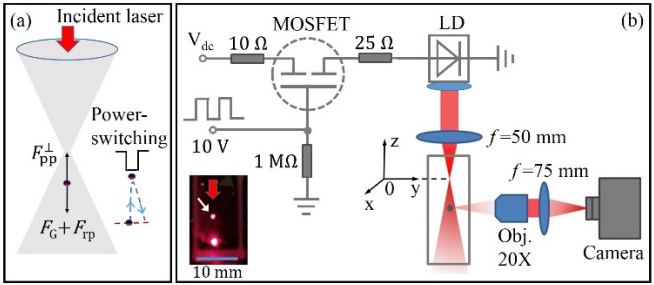

FIG. 1 (a) An absorbing particle is vertically trapped by a downward focused Gaussian beam and is held by an upward component of $F_{pp}$ to balance the other downward forces. A power-switching in incident laser causes dynamic motion of the trapped particle. (b) Schematic setup for observing the dynamic motion of a trapped particle with a high-speed camera. The power of laser diode (LD) can be rapidly modulated with a rise or fall time <1 μs via a pulsed driving current. The inset shows the image of a trapped particle.

### A. Optical trapping by negative photophoretic force

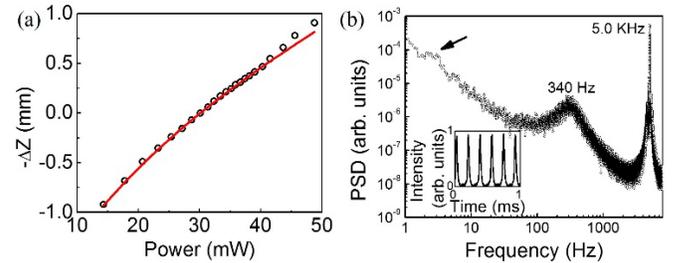

FIG. 2 (a) The position displacement of a trapped particle (Particle 1) vs the incident laser power. The red curve shows the theoretical fitting of the experimental data by Eq. (3) with the parameter $z_1$=3.0 mm at $P_1$=30.2 mW; (b) The power spectral density (PSD) of the Brownian motion of Particle 1 illuminated by the laser beam with a constant power of 30.2 mW. The natural frequency of potential well and the rotation frequency of particle's circular motion can be determined from the power spectrum. The corner frequency can also be determined as marked by an arrow. The insert shows the scattering intensity from the trapped particle, which represents the rotation motion of the particle at a rotation frequency of 5.0 kHz.

The focused laser is a Gaussian beam and the intensity distribution in particle's position can be determined by,

$$I_z = \frac{I_F}{1+(z/z_R)^2}, \quad (1)$$

where $I_F=P/(\pi w_0^2)$ is the intensity at the focal point ($z$=0) with the beam waist width of $w_0$, and $z_R=\pi w_0^2/\lambda$ is the Rayleigh range of the focused Gaussian beam. Since the trapping positions of the particle is far away from the focal point (such that $z \gg z_R$), Eq. (1) can be simplified as

$$z = z_R\sqrt{I_F/I_z}. \quad (2)$$

Assume that the laser intensity $I_z$ in the position of the trapped particle is kept the same when the incident power is changed, the position displacement ($\Delta z=z-z_1$) of the trapped particle can be calculated by

$$\Delta z = z_1(\sqrt{I_F/I_{F1}} - 1), \quad (3)$$

where $I_{F1}$ is the intensity at the focal point when the particle is trapped at $z=z_1$.

The theoretical curve from the Eq. (3) is drawn in FIG. 2(a) (the red curve). We found that the position displacement ($\Delta z=z-z_1$) of the trapped particle can be theoretically fitted very well with the value $z_1$=3.0 mm. This consistency indicates that the laser intensity in the trapping position tends to be the same when the incident power is changed, although the axial trapping position is changed. This finding suggests that when the incident power is changed, the variation of particle's temperature (and thus $F_{pp}$) may not change fast enough so that particle is pushed away or pulled back by the change in radiation pressure force, and the particle is thus moved to a new position where the actual light intensity on the particle is nearly the same. The Brownian motion in z axis direction of the trapped particle was measured with a position detector as shown in FIG. 2(b), from which the natural frequency (~340 Hz) of potential well and the corner frequency of the photophoretic trap can be determined [26], while the sharp peak at 5.0 kHz corresponds to the rotation frequency of the trapped particle around the optical axis [15].

## B. Dynamic motion of the trapped particles following laser intensity modulation

In order to measure the time dependence of negative $F_{pp}$, a particle (a glassy carbon sphere with a size ~4 μm) is first trapped with a constant intensity $I_0$ and then a low-intensity pulse is applied to modulate the laser power, followed by measuring the dynamic motion of the trapped particle. Figures 3(a) and 3(b) show that when the laser power is turned off, the particle is found to move upward for ~50 μm with a nearly constant speed. When the laser intensity returns to the original level, the particle is pushed back to the original equilibrium position in a longer time interval. Since $F_{rp}$ is quickly reduced during low-intensity pulse and $F_G$ is unchanged, the upward motion of the particle at a constant velocity indicates that the $F_{pp}$ force keeps unchanged for a short time and thus particle's temperature in this interval remains nearly unchanged. When the laser power is turned back, the light intensity on the up-shifted particle is higher than the intensity at the trapping position, causing greater $F_{rp}$ even the $F_{pp}$ force remains the same. Therefore, the particle is pushed back to the original trapping position when incident power is recovered after the low-intensity modulation pulse. Similar dynamic motion was observed for a trapped micron-sized cluster (with a size ~6 μm) of carbon nanoparticles when the laser intensity is switched to a low intensity (~5% of $I_0$ for the observation of particle's images), as shown in FIGs. 3(c) and 3(d). When the laser intensity is switched to a low intensity, the particle is found to move upward for ~190 μm with a nearly constant speed, suggesting that the $F_{pp}$ force keeps unchanged for a short time.

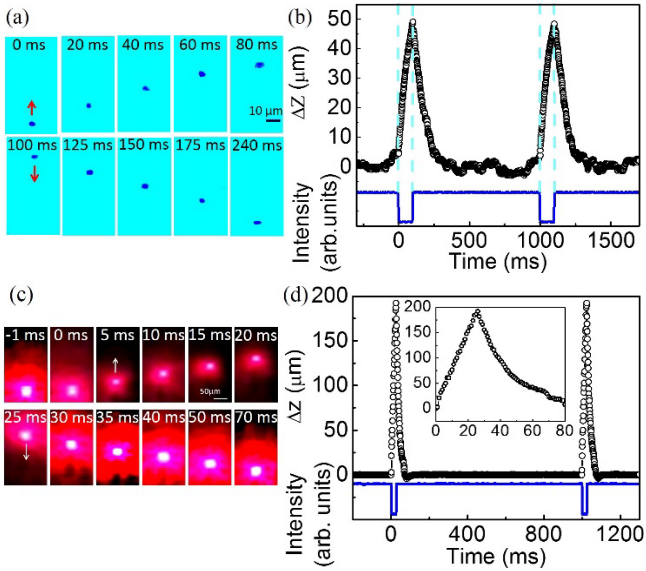

FIG. 3 (a) Sequential bright-field images of a trapped glassy carbon sphere (with a size ~4 μm) following the power-switching of 100 ms. (b) The variation of particle's position versus time. (c) Sequential images of a trapped micron-sized cluster (with a size ~6 μm) of carbon nanoparticles (Particle 1) undergoing the dynamic motion following the power-switching with a negative square pulse of 25 ms. (d) The variation of cluster's position versus time. The blue curve shows the change in light intensity, and the inset shows the magnified variation of particle's position after the application of a negative square pulse (see Supplemental Video 1 [29]).

More results on dynamic motion of the trapped particles following laser intensity modulation are shown in Supplemental Videos, in which Supplemental Video 1 (recorded by scattering light images) shows the dynamic motion of a trapped cluster of carbon nano-particles when the incident laser power is switched to a low-level for a square pulse duration of 25 ms [29]; Supplemental Video 2 (recorded by shadow images) shows the dynamic motion of a trapped particle when the incident laser power is switched to a low-level for a square pulse duration of 25 ms [29]; Supplemental Video 3 shows the dynamic motion of a trapped particle when the incident laser power is switched to a high-level for a square pulse duration of 5 ms [29].

### C. Temporal dependence of photophoretic force

The experimental results above clearly demonstrate that the temporal change in photophoretic force $F_{pp}$ (the sum of $F_{\Delta\alpha}$ and $F_{\Delta T}$ forces) in response to the intensity modulation is remarkably slow, which could be due to the slow change in particle's temperature. A semi-empirical expression for the $F_{\Delta\alpha}$ force is given by Rohatschek [9], for spherical particles whose surface is divided into two hemispheres with two different accommodation coefficients $\alpha_1$ and $\alpha_2$:

$$F_{\Delta\alpha} = \frac{1}{12\bar{c}} \frac{1}{1+(p/p^*)^2} \frac{\Delta\alpha}{\bar{\alpha}} H. \quad (4)$$

Here $\bar{c}$ is the mean speed of molecules, $p$ is the pressure, $p^*$ is the characteristic pressure, $\Delta\alpha=\alpha_1-\alpha_2$, $\bar{\alpha}=(\alpha_1+\alpha_2)/2$, and $H$ denotes the net energy flux transferred by the surrounding molecules. The ratio of gas mean free path $\lambda$ to particle diameter is defined as the Knudsen number $K_n$. $H$ can be written as $H_{cont}=4\pi a K_g(T_s-T_0)$ in the continuum-flow regime ($K_n \ll 1$), and $H_{free}=4\pi a^2 h(T_s-T_0)$ in the free-molecular regime ($K_n \gg 1$) [9,11]. Here $K_g$ is the thermal conductivity of surrounding gas, $h$ the convection heat transfer coefficient in still gas, $a$ the radius of the particle, $T_s$ and $T_0$ are the temperature of the particle's surface and the gas molecules, respectively. Consequently, the $\Delta\alpha$-force can be expressed by $F_{\Delta\alpha}=k_3(T_s-T_0)$, with a parameter $k_3$.

The rate of change of particle's heat energy can be written as:

$$\frac{dQ}{dt} = -H - E_{emi} + E_{abs} + AI, \quad (5)$$

where $Q=mc_V(T_s-T_0)$, $c_V$ is the specific heat of particles, $AI$ the power of the light absorbed by the particle with the absorption cross section $A$ and the incident irradiance $I$ at particle's position, and $E_{emi}$ and $E_{abs}$ the emitted and absorbed energy fluxes due to thermal radiation, respectively. Except for very low pressures, the net radiation flux $E_{emi}-E_{abs}$ is negligibly small compared to $H$ [9,10]. Accordingly, Eq. (5) can be rewritten as

$$\frac{d}{dt}(T_s - T_0) = -k_1(T_s - T_0) + k_2 AI, \quad (6)$$

where $k_1=4\pi a K_g/(mc_V)$ or $k_1=4\pi a^2 h/(mc_V)$ for the continuum-flow and the free-molecular regime, respectively, and $k_2=1/(mc_V)$.

Most of previous theoretical and experimental work considered thermal equilibrium with a constant light intensity [9-25], such that $H=AI$ from Eq. (5). Therefore, both the $\Delta\alpha$- and $\Delta T$- forces in steady state are directly proportional to the light intensity [9-12]. However, Eq. (6) indicates that the time constant of the change in particle's temperature as well as the $F_{\Delta\alpha}$ force is $\tau_1=1/k_1$. When the illumination light is turned off ($I=0$), they decay exponentially, $T_s-T_0=\Delta T_0 \exp(-k_1 t)$, and thus $F_{\Delta\alpha}=k_3(T_s-T_0)=F_{\Delta\alpha,0}\exp(-k_1 t)$, where $\Delta T_0$ and $F_{\Delta\alpha,0}$ are the temperature difference and photophoretic force (since $F_{\Delta T}$ is much smaller than $F_{\Delta\alpha}$ for the trapped particle, see Sec. III) in steady state before the change of intensity, respectively. For a micron-sized spherical particle or irregular aggregate of nano-particles, the value of the time constant $k_1^{-1}$ had not been measured. By modeling the dynamic motion of the trapped particle following a pulsed change in laser intensity, it is possible to measure the time constant of the change in photophoretic force $F_{pp}$ as well as in particle's temperature (see Sec. IV).

## III. THEORETICAL ESTIMATION OF THE PHOTOPHORETIC $F_{\Delta T}$ AND $F_{\Delta\alpha}$ FORCES

Wurm and Kraus [12] were the first to use a downward convergent laser beam to levitate micron-sized graphite aggregates (consisting of smaller flakes), which perform negative photophoresis. The question that which of the forces, $F_{\Delta T}$ or $F_{\Delta\alpha}$, is responsible for the particle levitation remains debate [9-11,18]. Wurm and Kraus suggested that negative $F_{\Delta T}$ force might arise for large complex aggregates if directed light can penetrate through pores into deeper layer of aggregate but cannot run away such that the backside could be hotter than the surface facing the illumination light [10,12], although they did not exclude the $F_{\Delta\alpha}$ force being responsible for the particle lift for low inclination of laser beam. This explanation with negative $F_{\Delta T}$ force is contradictory to Rubinowitz model of radiometric force [30]. According to Rubinowitz model, big agglomerate should perform positive photophoresis and not negative. The $\Delta T$-force is always longitudinal (either positive or negative) determined by the light beam because it is caused by inhomogeneous heating of the particle [10]. For weakly absorbing particles (such as glass spheres coated with a thin Au film [19]), the maximum absorption may be at the backside due to the lens effect, which leads to the backside hotter than the surface facing the illumination light. Therefore, a negative $F_{\Delta T}$ force could be generated for weakly absorbing particles. If the particle size is comparable to the wavelength of light (with the size parameter $2\pi a/\lambda=0.5$-5 or diameter $2a$ <0.8 μm for $\lambda=0.5$ μm), even strong absorbing particles (with refractive index $m=1.95-0.66i$ for carbon spheres) act as structured absorption and may perform negative photophoresis [31,32]. However, for strongly absorbing particles (such as graphite and carbon particles) that have a size well larger than the wavelength, there will always be more absorption on illuminated side, and therefore only positive photophoretic $F_{\Delta T}$ force is possible. For micron-sized carbon spheres, graphite powders or aggregates of carbon nano-particles with 2-10 μm in diameter used in our experiments, it is more likely that the $F_{\Delta T}$ force is positive. In particular, for glassy carbon spheres (2-12 μm) and natural graphite powders (~5 μm), there are not pores for the light penetrating into deeper layer of the particle. Therefore, Wurm and Kraus's suggestion of negative $F_{\Delta T}$ force due to pore structures for large complex nano-powder aggregates is not likely applied for the absorbing solid particles used in our experiment.

### A. Thermal conductivity parallel to the surface or perpendicular to the surface

The magnitude of photophoretic $F_{\Delta T}$ force is highly dependent on the thermal conductivity $k_p$ of the particles [9-11], see below in Eq. (8). Glassy carbon spherical powder (size of 2-12 μm) used in this experiment is made of glass-like carbon and has high thermal conductivity, high temperature resistance, and hardness [33]. From vendor's specifications [28], the density is $\rho_p=2.2$ g/cm$^3$ and thermal conductivity of the glassy carbon spherical powder is $k_p=119$-165 W/(m·K), which is significantly higher than "normal" thermal conductivity ($k_t=6.3$ W·m$^{-1}$·K$^{-1}$) of natural graphite that is for thru-thickness heat conduction [33]. The high thermal conductivity of glassy carbon spheres is probably due to the fast heat transfer parallel to the surface (with in-plane thermal conductivity $k_i$), rather than the slow heat conduction perpendicular to the surface through thickness of the carbon material (with thru-plane thermal conductivity $k_t$), as shown in FIG. 4. It is well known that crystalline graphite has a thermal conductivity of 1200 W/(m·K) parallel to the layer planes but has a low thermal conductivity of ~6 W/(m·K) perpendicular to the layer plane [11]. It has been demonstrated that a natural graphite sheet (with 0.25-0.5 mm thickness) has asymmetry thermal conductivities with $k_i$=140-500 W/(m·K) and $k_t$=3-10 W/(m·K), which has been widely applied for heat spreaders [33]. The good thermal conductivity $k_i$ parallel to the spherical surface considerably reduces the temperature difference between the illuminated side and the backside of the particle, and thus reduces the value of $F_{\Delta T}$ force.

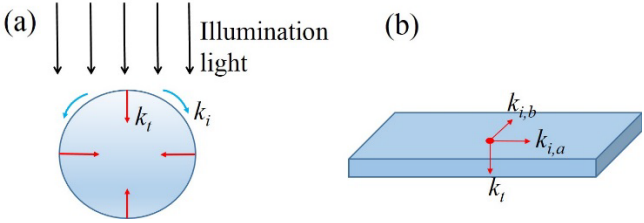

FIG. 4 In-plane thermal conductivity $k_i$ and thru-plane thermal conductivity $k_t$ of spherical surface (a) and planer surface (b).

## B. Calculation of photophoretic $F_{\Delta T}$ and $F_{\Delta\alpha}$ forces, gravitation force, radiation pressure force, and buoyant force for micron-sized glassy carbon spheres

A semi-empirical expression for photophoretic force is given by Rohatschek, in which under a thermal equilibrium condition both the $F_{\Delta\alpha}$ and $F_{\Delta T}$ forces are the function of gas pressure $p$ given by [9,11]

$$F_{\Delta\alpha} = \frac{1}{12\bar{c}} \frac{1}{1+(p/p^*)^2} \frac{\Delta\alpha}{\bar{\alpha}} \pi a^2 I(z), \quad (7)$$

and

$$F_{\Delta T} = \frac{2}{p/p_{max}+p_{max}/p} D \sqrt{\frac{\bar{\alpha}}{2}} \frac{a^2 J_1}{k_p} I(z), \quad (8)$$

where $k_p$ is thermal conductivity of the particle of radius $a$, $\bar{\alpha}$ is the average of two accommodation coefficients for two hemispheres of a spherical particle, $\Delta\alpha$ is the difference of the accommodation coefficients, $J_1$ is the asymmetry parameter of the particle for light absorption ($J_1$=0.5 for opaque particles, $J_1$ has the opposite sign if the absorption is mainly at the back side for negative photophoresis), and $p^*$ is a characteristic pressure:

$$p^* = \frac{1}{2a}\sqrt{3\pi\kappa}\bar{c}\eta, \quad (9)$$

$$p_{max} = \sqrt{\frac{2}{\bar{\alpha}}} p^*. \quad (10)$$

$D$ denotes a constant determined entirely by the state of the gas,

$$D = \frac{\pi}{2}\sqrt{\frac{\pi}{3}}\kappa\frac{\bar{c}\eta}{T_0}, \quad (11)$$

$$\bar{c} = \sqrt{\frac{8}{\pi}\frac{RT_0}{M}}, \quad (12)$$

where $\bar{c}$ is the mean speed of the gas molecules, $\kappa$ is a thermal creep coefficient, 0.9<$\kappa$<1.14 [9], $R$ is the universal gas constant, $M$ is the molar mass, $T_0$ is the temperature of the gas, and $\eta$ is the dynamic viscosity of the gas.

For micron-sized particles, the particle size is much larger than the gas mean free path of $\lambda$ = 66 nm at standard conditions ($p$=760 Torr, $T_0$=20 °C [11] and $p^*$=51.7 Torr for $d$=4 μm) and continuum-flow regime is valid at a high pressure ($p \gg p^*$). From Eqs. (7) and (8), the ratio of $F_{\Delta\alpha}$ and $F_{\Delta T}$ forces is inversely proportional to the pressure [9]. The particular pressure (at which the $F_{\Delta\alpha}$ and $F_{\Delta T}$ forces become equal) increases with thermal conductivity $k_p$ of the particle. For good thermal conductors, this pressure is far beyond $10^5$ Pa. This means that for atmospheric pressure or below, practically only the $F_{\Delta\alpha}$ force is dominant [10]. Here, we estimate the values of $F_{\Delta\alpha}$ and $F_{\Delta T}$ forces for glassy carbon spherical particles with the light intensity at the trapping position.

For the incident laser power of $P$=30.2 mW, wavelength $\lambda$=0.66 μm, and beam size of $D_0$=2.6 mm focused by a lens of $f$=50 mm, the beam waist width at the focal point is $w_0$=2$f\lambda$/($\pi D_0$)≈8.0 μm. The intensity at the focal point ($z$=0) is $I_F$=$P$/($\pi w_0^2$)≈15 kW/cm$^2$. The Rayleigh range is $z_R$=$\pi w_0^2$/$\lambda$≈0.3 mm. The light intensity at the trapping position of $z$=2 or 3 mm is given by $I$=330 W/cm$^2$ or 150 W/cm$^2$ calculated by $I_z$=$I_F$/[1+($z$/$z_R$)$^2$].

For a trapped glassy carbon spherical particle with a diameter $d$=4 μm, the mass is $m$=7.37×10$^{-14}$ kg. Gravitation force is $F_G$=0.722 pN. Buoyant force at the atmospheric pressure ($p$=760 Torr at 20 °C) is $F_B$=4×10$^{-4}$ pN, which is 3 orders of magnitude smaller than $F_G$ and thus can be neglected, because the density of the air is much less than the density of the solid carbon particles [18]. Radiation pressure force acting on the sphere is $F_{rp}$=($\pi a^2$)$I$/$c$=0.062 pN=0.086 $F_G$ for $I$=150 W/cm$^2$, which is an order of magnitude smaller than $F_G$. As calculated by Eqs. (7) and (8), FIG. 5 shows the values of $F_{\Delta T}$ and $F_{\Delta\alpha}$ forces as the function of pressure for various particle size with $k_p$=6.3W/(m·K) and $k_p$=120 W/(m·K) illuminated by light intensity of 150 W/cm$^2$, respectively. For 4-μm glassy carbon spheres ($k_p$=120 W·m$^{-1}$·K$^{-1}$) at $p$=760 Torr, $F_{\Delta\alpha}/F_G$=3.1, but $F_{\Delta T}/F_G$=0.23, indicating that $F_{\Delta T}$ force is an order of magnitude smaller than $F_{\Delta\alpha}$ force. This indicates that $F_{\Delta T}$ force alone is not sufficient to lift the sphere. When the pressure is decreased, the ratio of $F_{\Delta\alpha}$ force over $F_{\Delta T}$ force increases. Similar results are valid for $d$=2 μm and 6 μm particles. Therefore, these results are consistent with the previous statement that $F_{\Delta\alpha}$ force is practically dominant for micron-sized glassy carbon spheres due to their good conductivity for atmospheric pressure or below [10].

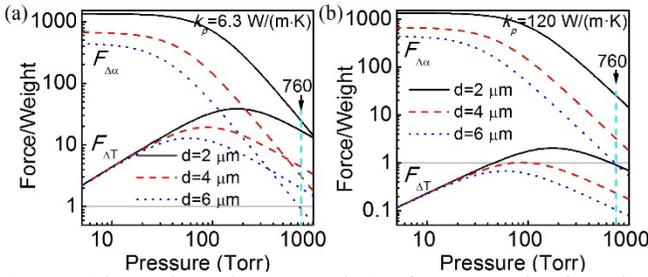

FIG. 5 Photophoretic $F_{\Delta T}$ and $F_{\Delta\alpha}$ forces as the function of pressure for various particle size. (a) Thermal conductivity $k_p$=6.3 W/(m·K) for amorphous graphite particles; (b) $k_p$=120 W/(m·K) for glassy carbon spherical particles. The following parameters are used for the calculation: $I$=150 W/cm$^2$, $\rho_p$=2.2×10$^3$ kg/m$^3$, $\bar{\alpha}$=0.7, $\Delta\alpha$=0.1, $J_1$=0.5, viscosity of air $\eta$=1.8×10$^{-5}$ Pa·s, and $\kappa$=1.14.

## IV. MODELING DYNAMIC MOTION AND TIME DEPENDENCE OF PHOTOPHORETIC FORCE

To model the dynamic motion of the trapped particle, we express the equation of motion in z direction as follow:

$$m\ddot{z} + \gamma\dot{z} = F_{pp} - F_{rp} - F_G + F_\xi(t)$$
$$= k_3(T_s - T_0) - AI/c - mg + F_\xi(t), \quad (13)$$

where $\gamma=6\pi\eta a$ is Stokes' drag constant, $\eta$ the viscosity of air, $a$ the radius of a particle assuming it is a sphere, $m$ the mass, $c$ the speed of light, the first term in the right side is $F_{\Delta\alpha}$, the second term the radiation pressure force, the third term $F_G$, and $F_\xi(t)$ the random force for Brownian motion. Since the range of dynamic motion due to intensity modulation is much larger than that of the Brownian motion, $F_\xi(t)$ is very small compared with the other forces and can be neglected. In Eqs. (6) and (13), the intensity $I(z,t)$ is the function of position $z$ (due to focusing of Gaussian beam) and time $t$ (due to power modulation). When illuminated with a constant intensity, Eqs. (6) and (13) reaches a steady state such that $F_{pp}=F_G+F_{rp}$, and the particle undergoes a circular motion [15]. When the laser intensity is switched to a low level, the radiation pressure $F_{rp}$ is reduced immediately and the particle undergoes a vertically upward motion (Supplemental Video 1), indicating that the $F_{\Delta\alpha}$ force is likely aligned to upright direction under the action of the gravitation force $F_G$.

Numerical simulation of Eqs. (6) and (13) could lead to comparison between experiment and theoretical modeling of dynamic motion of the trapped particle. Figure 6(a) shows experimental data of the displacement of a trapped particle (with a size ~6 μm) when a low-intensity square pulse or a high-intensity square pulse is applied, while the low-intensity level is ~5% of $I_0$ and the high-intensity level is ~2$I_0$. Supplemental Video 2 (bright-field image) and Supplemental Video 3 (scattering light image) show the dynamic motions of this particle after the application of the low-intensity and high-intensity pulses, respectively. As predicted, a high-intensity square pulse causes the particle be pushed downward, whereas a low-intensity square pulse causes the particle be pulled upward. Figure 6(b) shows numerical simulation of particle's displacement with similar low-intensity or high-intensity pulse modulation and with a simulation parameter $k_1$~1 s$^{-1}$. The theoretical calculations are consistent with the experimental results.

The temperature change of microparticles determines the time-dependence of $F_{pp}$ force. The results in FIGs. 3 and 6(a) show that $F_{\Delta\alpha}$ and the particle's temperature have no observable change during a short interval (25-ms for a trapped cluster of carbon nano-powders and 100-ms for a trapped glassy carbon sphere) of the intensity modulation. In order to find how long it will take for the particle's temperature to have a significant change after the power-modulation, we extended the duration of the low-intensity square pulse to 100-250 ms and expected to observe nonlinear change in particle's displacement due to the decrease in $F_{\Delta\alpha}$ when the laser power is turned off for a longer interval.

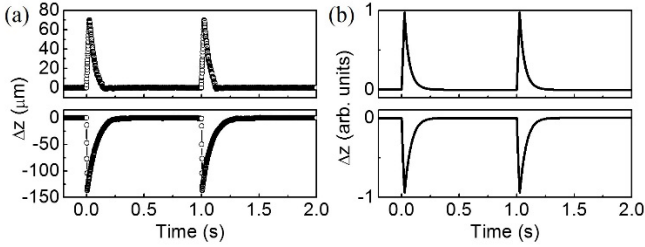

FIG. 6 (a) Experimental data of the position variation of a trapped particle (with a size ~6 μm) in z axis direction versus time after the application of a low-intensity square pulse of 25 ms (above) and a high-intensity square pulse of 5 ms (below). (b) Theoretical modeling of particle's displacement with a low-intensity square pulse (above) and a high-intensity square pulse (below). The simulation parameter for $k_1$ is ~ 1 s$^{-1}$.

Figures 7(a) and 7(b) show the position of a glassy carbon sphere and a cluster of carbon nano-powders as the function of time after the low-intensity modulation, respectively, with theoretical fitting. In the first period (~ 40 ms), the velocity of the particle is nearly constant, consistent with FIGs. 3 and 6(a). In the late period, the velocity of the particle declines slightly, indicating the magnitude of $F_{\Delta\alpha}$ force slightly decreases.

Ignoring the random force $F_\xi(t)$ and the light intensity of the low-intensity square pulse, we can obtain the particle's position and velocity in z axis as the function of time from Eqs. (6) and (13) under large viscosity condition where $m/\gamma$ is much smaller than the time interval of measurement:

$$z = \frac{1}{\gamma} F_{\Delta\alpha,0} \cdot \tau_1 \cdot (1 - e^{-t/\tau_1}) - \frac{mg}{\gamma} t, \qquad (14)$$

$$v = \frac{1}{\gamma} F_{\Delta\alpha,0} \cdot e^{-t/\tau_1} - \frac{mg}{\gamma}, \qquad (15)$$

where $\tau_1 = 1/k_1$ is the time constant of temperature change of the trapped particle and $F_{\Delta\alpha,0}$ is the magnitude of the $F_{\Delta\alpha}$ force before intensity modulation. The experimental data were fitted very well with Eq. (14) as shown in a red curve in FIG. 7, with the parameter $\tau_1$ equal to 0.5 s and 0.1 s for the nano-powder cluster and for glassy carbon sphere, respectively. Due to the non-symmetry between the upward and downward motions during a square-pulse modulation, we also find that the trapping position of the particle may vary with the modulation frequency with the same average power (see FIG. 11 in Appendix B).

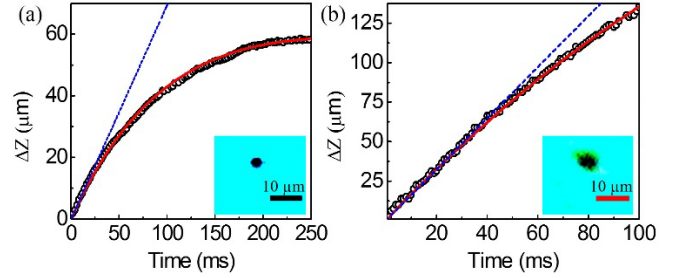

FIG. 7 (a) The position of a trapped glassy carbon sphere (open circles) vs time when the laser power is switched off for 250 ms. The blue dashed line is a line with the slope equal to the average velocity in the first period. The red solid curve is the theoretical fitting with $\tau_1$ equal to 0.1 s. The inset is the image of the trapped particle. (b) The position of a trapped cluster of carbon nano-powders (open circles) vs time when the laser power is switched off for 100 ms. The blue dashed line is a line with the slope equal to the average velocity in the first period. The red solid curve is the theoretical fitting with $\tau_1$ equal to 0.5 s. The inset is the image of the trapped cluster.

## V. CHARACTERIZING THE PROPERTIES OF TRAPPED PARTICLES BY MEASURING THEIR DYNAMIC MOTION FOLLOWING THE SWITCHING OF LASER POWER

Since the photophoretic force $F_{\Delta\alpha}$ is dependent on the properties of particles and ambient gases, we may obtain some physical properties of trapped particles by measuring their dynamic behaviors under the power modulation.

According to Eq. (15), the ratios of the $F_{\Delta\alpha}$ force and gravitation force to the drag constant, $F_{\Delta\alpha,0}/\gamma$ and $mg/\gamma$, satisfy the following relationship at $t=0$:

$$v_0 = \frac{F_{\Delta\alpha,0}}{\gamma} - \frac{mg}{\gamma}, \qquad (16)$$

where $v_0$ is the average velocity of the particle in the first ~20 milliseconds after the power is switched off, in which the particle nearly moves at a constant speed. By extending the duration of the low-intensity square pulse to 500 ms, we can measure the position curve of a trapped glassy carbon sphere as the function of time, as shown in FIG. 8. The experimental data can be very well fitted by Eq. (14) as the red curve in FIG. 8 with the following parameters:

$$\frac{mg}{\gamma} = 0.18 \text{ mm/s}, \qquad (17)$$

$$\frac{F_{\Delta\alpha,0}}{\gamma} = 1.19 \text{ mm/s}, \qquad (18)$$

$$\tau_1 = 101.5 \text{ ms}, \qquad (19)$$

where $\tau_1$ is the time constant of temperature change of the glassy carbon sphere with $v_0 = 1.01$ mm/s. From Eq. (17), we may obtain $m/\gamma = 1.84 \times 10^{-5}$ s for the trapped glassy carbon sphere, which represents the ratio of the mass to the drag constant of the particle in air. Due to the small size of the particles, these quantities are difficult to be directly measured. Our experiment provides a potential approach to determine these physical quantities of the particles in air.

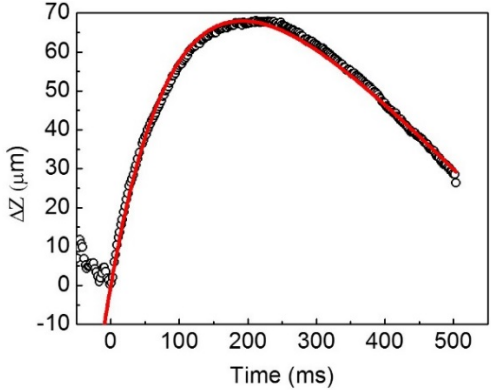

FIG. 8 The position of a trapped glassy carbon sphere (open circles) vs time when the laser power is switched off for 500 ms. The red solid curve is the theoretical fitting with $\tau_1$ approximately equal to 0.1 s. The inset is the image of the trapped particle.

We may compare the measured value of $m/\gamma$ with the value determined by another method developed by Bera et al. [25]. Following their approach, the motion of the trapped particle in the axial direction is detected by a position sensitive detector at a constant laser power of 30.2 mW. We analyze the data in the frequency domain by determining the power spectral density (PSD) of the Brownian motion and the result is shown in FIG. 9. The PSD is given by [25]

$$S(\omega) = \beta^2 \frac{2 k_B T}{k} \frac{\Omega^2 \Gamma}{(\Omega^2 - \omega^2)^2 + \omega^2 \Gamma^2}. \quad (20)$$

Here, $\beta^2$ is the conversion factor of the detector from voltage to actual displacement, $k_B$ is the Boltzmann constant, $T$ is the temperature, $k$ is the stiffness, and $\Omega$ and $\Gamma$ are defined as $\Omega^2 = k/m$ and $\Gamma = \gamma/m$, respectively. The PSD data can be theoretically fitted with Eq. (20) by taking the fitting parameters as $2 k_B T \beta^2/k = 4 \times 10^{-5}$ s$^2$, $\Omega = 900$ s$^{-1/2}$, $\gamma/m = 54000$ s$^{-1}$, as shown with red curve in FIG. 9. This leads to the value of $m/\gamma = 1.85 \times 10^{-5}$ s, which is very close to the value determined by the dynamic motion following power modulation. Given the known values of the viscosity of air and the size of particles, the mass of the trapped particles can be determined [25]. For the glassy carbon particle with $d=2$ μm and the air viscosity $\eta = 1.8 \times 10^{-5}$ Pa·s, the mass of the particle is determined as $m = 6.3 \times 10^{-15}$ kg, which is slightly smaller than the theoretical value of $9.2 \times 10^{-15}$ kg assuming the sphere's bulk density of 2.2 g/cm$^3$. It should be noted that the effective density of the micro-sized particles might be less than the density of the bulk material.

In addition, from the fitting parameters of our approach, we may also determine the magnitude of photophoretic force in the scale of the gravity of the trapped particles, $F_{\Delta\alpha,0} = 6.6 mg$, and its time constant $\tau_1$ for the temperature change. To the best of our knowledge, it is the first direct characterization of photophoretic force and its temporal property.

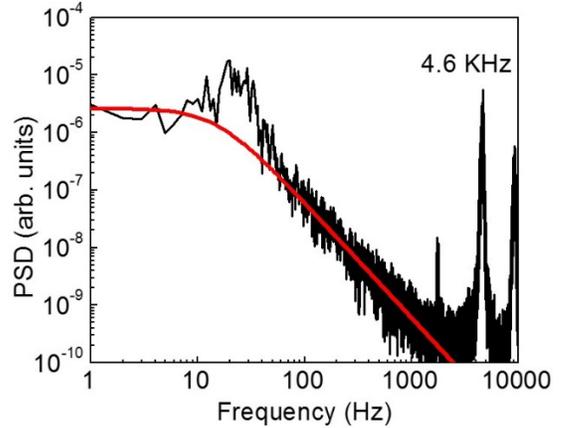

FIG. 9 The power spectral density (PSD) of the Brownian motion of a trapped glassy carbon sphere illuminated by the laser beam with a constant power of 30.2 mW. The natural frequency of potential well (~4.6 KHz) can be determined from the power spectrum.

## VI. CONCLUSIONS

In conclusion, we have demonstrated that the time dependence of optically-induced photophoretic force on absorbing particles by a power-modulated laser could be very slow due to the slow change in particle's temperature. A theoretical model has been developed to calculate the temperature and the dynamic motion of trapped particle, and well explain the experimental observations on the particle's position in response to the power modulation. The ultimate mechanism of negative photophoretic force is the difference in thermal accommodation coefficient on the top (laser illuminated) and bottom sides of the particles. Although photophoretic force under constant illumination is well understood, how this force temporally responds to the fast change in light intensity is unclear and has rarely been studied. The novelty reported here is the long time-scale of the effect, which is of the order of seconds. Our work could be of interest for applications of particle manipulations including the control of photophoretic-based optical pulling, transportation and manipulation of atmospheric particles. We also demonstrate that some physical properties of the trapped light-absorbing particles (such as mass, temperature change, or gas viscosity) could be characterized by measuring their dynamic motion following the switching of laser power.

## SUPPORTING INFORMATION

Additional supporting information may be found in the online version of this article at the publisher's website.

Video S1: A movie (scattering light images) showing the dynamic motion of a trapped cluster (Particle 1) of carbon nano-powders when the incident laser power is switched to a low-level for a square pulse duration of 25 ms.

Video S2: A movie (shadow images) showing the dynamic motion of a trapped particle (Particle 2) when the incident laser power is switched to a low-level for a square pulse duration of 25 ms.

Video S3: A movie (scattering light images) showing the dynamic motion of a trapped particle (Particle 2) when the incident laser power is switched to a high-level for a square pulse duration of 5 ms.

## ACKNOWLEDGMENTS

This work was supported by the National Natural Science Foundation of China (Grant No. 61775036) and by the Scientific Research Foundation for High-level Talents (Innovation Team) of Dongguan University of Technology (Grant. No. KCYCXPT2017003). *Corresponding author: liy@ecu.edu

## APPENDIX A: MATERIALS OF LIGHT-ABSORBING PARTICES USED IN THE TRAPPING EXPERIMENT

The light-absorbing particles of different sizes used in this experiment were made of carbon materials: (1) glassy carbon spherical powders (size of 2-12 µm, Sigma-Aldrich, item#: 484164-10G); (2) carbon nano-powder clusters (size<50 nm, Sigma-Aldrich, item#: 633100-25G); (3) natural graphite powders (size ~5 µm, US Research Nanomaterials, US1158M); and (4) natural graphite nano-powders (size of 400-1200 nm, US Research Nanomaterials, item#: US1058). Figure 10 shows the images of these particles observed with a microscope. Because of strong absorption in visible wavelengths (400-700 nm), all these micro-sized particles appear as dark under the microscope.

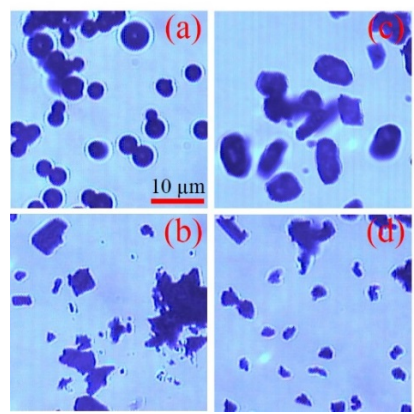

FIG. 10 Microscopic images of light-absorbing particles used in the experiment. (a) Glassy carbon spherical powders (size of 2-12 µm); (b) Carbon nano-powder (size <50 nm) clusters; (c) Natural graphite powders (size ~5 µm); (d) Natural graphite nano-powders (size of 400-1200 nm).

## APPENDIX B: DEPENDENCE OF THE TRAPPING POSITION ON THE MODULATION FREQUENCY OF A POWER-MODULATED LASER BEAM

The inset in Fig. 11(b) shows the positions of a particle trapped by power-modulated laser beam with frequency equal to 100 KHz and 100 Hz, respectively. It can be seen clearly that the particle is pushed away with low modulation frequency. A given particle tends to have two stable locations with high and low frequency modulation. Under high modulation frequency (>50 kHz), the particle's trapping position is the same as that with the continuous-wave irradiation of the same average power, since the particle cannot follow such a high modulation frequency to move its position due to air viscosity. However, under low modulation frequency (<100 Hz), the particle tends to move to the position irradiated by the continuous-wave laser beam with twice the power. To understand this result, it should be noted that one cycle of the square-modulation light consists of half cycle with zero intensity and half cycle with twice of the average intensity $I_0$, see Fig. 11(a). During the half cycle of $2I_0$, the particle is pushed forward by the radiation pressure. During the half cycle with zero intensity, the particle is pulled back. However, the displacement in latter process is much slower than that in former process. After multiple cycles, the particle is pushed closer to the position of the particle trapped by the continuous wave laser beam with twice the power.

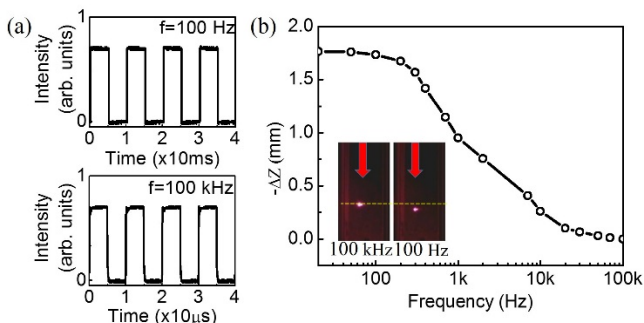

**FIG. 11** (a) The intensity of trapping beam is modulated by a square function with a frequency equal to 100 Hz or 100 kHz. (b) The position shift of the trapped particle at different modulation frequencies of the square-function modulated beam. The inset shows the images of the particle trapped by a power-modulated beam with the frequency of 100 Hz and 100 kHz, respectively. It can be seen clearly that the particle is pushed away under modulated light with low frequency.

**Keywords**: photophoretic force, optical trapping, temporal response, absorbing particles, intensity modulation

**References:**

[1] A. Ashkin, Acceleration and trapping of particles by radiation pressure, Phys. Rev. Lett. **24**, 156 (1970).
[2] D. G. Grier, A revolution in optical manipulation, Nature **424**, 810 (2003).
[3] M. A. Taylor, M. Waleed, A. B. Stilgoe, H. Rubinsztein-Dunlop, and W. P. Bowen, Enhanced optical trapping via structured scattering, Nat. Photonics **9**, 669 (2015).
[4] T. Li, S. Kheifets, D. Medellin, M. G. Raizen, Measurement of the instantaneous velocity of a Brownian particle, Science **328**, 1673 (2010).
[5] D. B. Ruffner and D. G. Grier, Optical conveyors: A class of active tractor beams, Phys. Rev. Lett. **109**, 163903 (2012).
[6] M. I. Petrov, S. V. Sukhov, A. A. Bogdanov, A. S. Shalin, and A. Dogariu, Surface plasmon polariton assisted optical pulling force, Laser Photonics Rev. **10**, 116 (2016).
[7] J. Gieseler, B. Deutsch, R. Quidant, and L. Novotny, Subkelvin parametric feedback cooling of a laser-trapped nanoparticle, Phys. Rev. Lett. **109**, 103603 (2012).
[8] T. Li, S. Kheifets, and M. G. Raizen, Millikelvin cooling of an optically trapped microsphere in vacuum, Nat. Phys. **7**, 527 (2011).
[9] H. Rohatschek, Semi-empirical model of photophoretic forces for the entire range of pressures, J. Aerosol. Sci. **26,** 717 (1995).
[10] O. Jovanovic, Photophoresis-light induced motion of particles suspended in gas, J. Quant. Spectrosc. Radiat. Transfer **110**, 889 (2009).
[11] H. Horvath, Photophoresis-a forgotten force?, KONA Powder Part. J. **31,** 181 (2014).
[12] G. Wurm and O. Krauss, Experiments on negative photophoresis and application to the atmosphere, Atmos. Environ. **42**, 2682 (2008).
[13] G. Wurm and O. Krauss, Dust eruptions by photophoresis and solid state greenhouse effects, Phys. Rev. Lett. **96,** 134301 (2006).
[14] V. G. Shvedov, A. V. Rode, Y. V. Izdebskaya, A. S. Desyatnikov, W. Krolikowski, and Y. S. Kivshar, Giant optical manipulation, Phys. Rev. Lett. **105,**118103 (2010).
[15] J. Lin, and Y. Q. Li, Optical trapping and rotation of airborne absorbing particles with a single focused laser beam, Appl. Phys. Lett. **104,** 101909 (2014).
[16] A. S. Desyatnikov, V. G. Shvedov, A. V. Rode, W. Krolikowski, and Y.S. Kivshar, Photophoretic manipulation of absorbing aerosol particles with vortex beams: theory versus experiment, Opt. Express **17**, 8201 (2009).
[17] Y.-L. Pan, C. Wang, S. C. Hill, M. Coleman, L. A. Beresnev, and J. L. Santarpia, Trapping of individual airborne absorbing particles using a counter flow nozzle and photophoretic trap for continuous sampling and analysis, Appl. Phys. Lett. **104**, 113507 (2014).
[18] N. Eckerskorn, R. Bowman, R. A. Kirian, S. Awel, M. Wiedorn, J. Küpper, M. J. Padgett, H. N. Chapman, and A. V. Rode, Optically induced forces imposed in an optical funnel on a stream of particles in air or vacuum, Phys. Rev. Appl. **4**, 064001 (2015).
[19] V. G. Shvedov, A. R. Davoyan, C. Hnatovsky, N. Engheta, and W. Krolikowski, A long-range polarization-controlled optical tractor beam, Nat. Photonics **8**, 846 (2014).
[20] J. Lin, A. G. Hart, and Y. Q. Li, Optical pulling of airborne absorbing particles and smut spores over a

meter-scale distance with negative photophoretic force, Appl. Phys. Lett. **106**, 171906 (2015).

[21] C. Wang, Z. Gong, Y.-L. Pan, and G. Videen, Laser pushing or pulling of absorbing airborne particles, Appl. Phys. Lett. **109**, 011905 (2016).

[22] J. Lu, H. Yang, L. Zhou, Y. Yang, S. Luo, Q. Li, and M. Qiu, Light-induced pulling and pushing by the synergic effect of optical force and photophoretic force, Phys. Rev. Lett. **118**, 043601 (2017).

[23] L. Ling and Y. Q. Li, Measurement of Raman spectra of single airborne absorbing particles trapped by a single laser beam. Opt. Lett. **38,** 416 (2013).

[24] C. Wang, Z. Gong, Y.-L. Pan, and G. Videen, Optical trap-cavity ringdown spectroscopy (OT-CRDS) as a single-aerosol-particle-scope, Appl. Phys. Lett. **107**, 241903 (2015).

[25] S. K. Bera, A. Kumar, S. Sil, T. K. Saha, T. Saha, and A. Banerjee, Simultaneous measurement of mass and rotation of trapped absorbing particles in air, Opt. Lett. **41**, 4356 (2016).

[26] D. E. Smalley, E. Nygaard, K. Squire, J. Van Wagoner, J. Rasmussen, S. Gneiting, K. Qaderi, J. Goodsell, W. Rogers, M. Lindsey, K. Costner, A. Monk, M. Pearson, B. Haymore and J. Peatross, A photophoretic-trap volumetric display, Nature **553**, 486 (2018).

[27] B. Hadad, S. Froim, H. Nagar, T. Admon, Y. Eliezer, Y. Roichman, and A. Bahabad, Particle trapping and conveying using an optical Archimedes' screw, Optica **5**, 551 (2018).

[28] Specification of glassy carbon spherical powder. Online available: https://www.nanoshel.com/product/glassy-carbon-spherical-powder/.

[29] See Supplemental Material at [URL will be inserted by publisher] for the dynamic motion of the trapped particles following laser intensity modulation.

[30] A. Rubinowitz, Radiometerkräfte und Ehrenhaftsche Photophorese (I and II), Ann. Phys. **62**(4), 691-715, 716-737 (1920).

[31] P. W. Dusel, M. Kerker, and D. D. Cooke, Distribution of absorption centers within irradiated spheres, J. Opt. Soc. Am. **69**, 55-59 (1979).

[32] W.-K. Li, P.-Y. Tzeng, C.-Y. Soong, and C.-H. Liu, Absorption center of photophoresis within micro-sized and spheroidal particles in a gaseous medium, Int. J. Elect. Comp. Ener. Electron. Comm. Eng. **4**(5), 803-807 (2010).

[33] M. Smalc, G. Shives, G. Chen, S. Guggari, J. Norley, R. A. Reynolds III, Thermal performance of natural graphite heat spreaders, Proceedings of IPACK2005, IPACK2005-73073, pp. 79-89, July 17-22, San Francisco, California USA. (Online available: http://citeseerx.ist.psu.edu/viewdoc/download?doi=10.1.1.736.9349&rep=rep1&type=pdf)